\documentclass[review]{elsarticle}

\usepackage{subfig}
\usepackage{dcolumn}
\usepackage{bm}
\usepackage{amsmath}
\usepackage{amssymb}
\usepackage{amsfonts}
\usepackage{amsbsy}
\usepackage{eucal}
\usepackage[T1]{fontenc}
\usepackage{multirow}
\usepackage{txfonts}
\usepackage{color}
\usepackage[normalem]{ulem}
\biboptions{sort&compress}

\renewcommand{\u}{\mathbf{u}}
\newcommand{\x}{\bm{x}}

\newcommand{\vx}{\mathbf{x}}
\renewcommand{\v}{\mathbf{v}}

\newcommand{\phib}{{\boldsymbol{\phi}}}

\newcommand{\alphab}{{\boldsymbol{\alpha}}}

\newcommand{\Bc}{\mathcal{B}}

\newcommand{\Oc}{\mathcal{O}}
\newcommand{\Uc}{\mathcal{U}}

\newcommand{\Nbb}{\mathbb{N}}

\newcommand{\Rbb}{\mathbb{R}}

\newcommand{\U}{{\mathbf{U}}}
\newcommand{\G}{{\mathbf{G}}}
\newcommand{\dd}{\mathrm{d}}

\newcommand{\eqeqref}{Eq. \eqref}

 \newcommand{\hadd}[1]{\textcolor{black}{#1}}

\DeclareSymbolFontAlphabet{\mathcal}{symbols}

\begin{document}

\begin{frontmatter}

\title{Scalable approximation of Green's function for estimation of anharmonic energy corrections}

\author{P. Rai, K.Sargsyan, H. Najm}
\address{Sandia National Laboratories, Livermore, California, 94551, USA}
\fntext[myfootnote]{Corresponding author: pmrai@sandia.gov}

\author{So Hirata}
\address{Department of Chemistry, University of Illinois at \\ Urbana-Champaign, Urbana, Illinois, 61801, USA}

%
%

\begin{abstract}
In \cite{RAI19}, we presented a method based on separated integration to estimate anharmonic corrections to energy and vibration of molecules in a second-order diagrammatic vibrational many-body Green's function formalism. A severe bottleneck in extending this approach to bigger molecules is that the storage of the Green's function scales exponentially with the number of atoms in the molecule. In this article, we present a method that overcomes this limitation by approximating the Green's function in the Hierarchical Tucker tensor format. We illustrate that the storage cost is linear in dimension and hence one can obtain accurate representations of the Green's function for a molecule of any size. Application of this method to estimate the second-order correction to energy of molecules illustrates the advantage of this approach.
\end{abstract}
\end{frontmatter}


\section{Introduction}

In quantum chemistry, accurate estimation of energy and vibrational frequencies of molecules requires integration of
functions whose dimensionality increases linearly with the number of nuclei in the molecule. In~\cite{RAI19} and~\cite{RAI17}, we introduced and extended a separated integration formulation for estimation of anharmonic energy and vibrational corrections  of molecules in XVH2. Application of this method for bigger molecules requires two necessary conditions. The first condition is to obtain an accurate approximation of the potential energy surface (PES) with as few evaluations as possible. A rigorous mathematical foundation for efficient representation of the PES can be found in the early works of Mezey~\cite{mezey1,mezey2,mezey3,mezey4,mezey5,mezey6,mezey7}. Studies from Dawes' group~\cite{Dawes1,Dawes2,Dawes3} focus on accurate fitting of the global PES. A many-body expansion of the PES, using such methods as the $n$-mode representation introduced by Bowman and coworkers~\cite{Nmodeapprox, BowmanPES2,BowmanPES,BowmanPES3}, addresses the curse of dimensionality, and has been extended by others \cite{Nmodeapprox2, Nmodeapprox3,Rabitz1} to problems other than PES representation~\cite{Rabitz2,Rabitz3}. Sum-of-products representations of PES, which serve as the basis of tensor decomposition, were  explored by J\"{a}ckle and Meyer~\cite{Jackle:1996, Jackle:1998}, proposing the {\sc potfit} and multilayer {\sc potfit} methods, and later by Otto~\cite{Otto:2014}, Carrington and coworkers~\cite{Manzhos:2006,Carrington}, and Ziegler and Rauhut~\cite{Rauhut,Rauhut2}. In our recent work~\cite{RAI19,RAI17}, we proposed techniques that exploit special mathematical structure (e.g. low rank, sparsity) of these functions.

The second condition, specific to second-order energy and frequency corrections in XVH2, is the requirement of efficient low rank approximation of the Green's function. A key bottleneck here is that the dimensionality of both the PES and the Green's function increases as $(m=3a - 6)$, where $a$ is the number of atoms. From an approximation point of view, this leads to an exponential increase in the number of multidimensional basis functions, given by $n^m$, where $n$ is the number of basis functions in each dimension, thus leading to difficulties in their storage and efficient approximation. This work targets this second requirement, i.e. efficient storage, approximation and application of Green's function to enable separated integration of anharmonic energy corrections of molecules.

There has been considerable work on the approximation of Green's function~\cite{hackbusch2008, hackbusch2008exp, beylkin2009,beylkin2005}, specifically in quantum chemistry~\cite{harrison04,khoromskij08}, most of which focuses on theoretical or numerical analysis of the approximation. In this work, however, we are concerned with a particular application of estimating anharmonic energy corrections. We therefore analyze the approximation of Green's function from the point of view of its effect on \hadd{the} accuracy of specific quantities of interest, in this case, the second\hadd{-}order correction to energy. In order to do so, we are interested in an approach that satisfies three conditions. Firstly, the approximation should have a polynomial representation. This is required because it enables efficient quadrature rules for numerical integration. Secondly, to reduce the computational cost, the approximation should have a low rank representation (see section \ref{sec:sepInt} below). Finally, for scalability considerations, we need a method to efficiently store the Green's function with a manageable storage cost and an approximation strategy whose computational cost at any point does not increase exponentially.

In this work, we satisfy these three conditions by proposing to store and approximate \hadd{the} Green's function in \hadd{the} Hierarchical Tucker tensor format \cite{Grasdyck:2010, Hackbusch:2009}, an efficient structured tensor format based on recursive subspace factorizations. The Hierarchical Tucker format is a specialization of the Tucker format and it contains canonical tensors as a special case \cite{Grasedyck:2013}. This format is a storage-efficient scheme to approximate and represent tensors which can be applied particularly well for approximation of high\hadd{-}order Green's function\hadd{s}. As will be seen in Section~\ref{sec:hierarchicalG}, this format permits storage and approximation of \hadd{the} Green's function with a complexity that grows only linearly with dimension, leading to a crucial efficiency improvement in estimating second\hadd{-}order anharmonic energy corrections in XVH2.

The outline of the paper is as follows. In Section~\ref{sec:sepInt}, we briefly recall \hadd{the} Green's function in \hadd{the} separated integration formulation for estimating second\hadd{-}order energy correction\hadd{s in} XVH2. Then, in Section~\ref{sec:lowrankG}, we discuss \hadd{the} tensor representation of \hadd{the} Green's function, and its approximation in several low\hadd{-}rank tensor formats, from a conceptual point view. In Section~\ref{sec:hierarchicalG}, we present and illustrate efficient storage of \hadd{the} Green's function in \hadd{the} Hierarchical Tucker tensor format. In Section~\ref{sec:anharmonicG}, we apply and illustrate the proposed method to estimate second\hadd{-}order energy corrections of select molecules, and derive conclusions in Section~\ref{sec:conclusion}.

\section{Green's function in XVH2}
\label{sec:sepInt}

As indicated in the previous section, efficient approximation of \hadd{the} Green's function in \hadd{the} Hierarchical Tucker format is pertinent to \hadd{the} second\hadd{-}order correction to \hadd{the} energy in the XVH2 formalism of quantum chemistry. The reader is referred to the original papers \cite{Hermes:2013,Hermes:2014} for the derivation of this formalism. Here, we briefly outline the method of separated integration~\cite{RAI17} to evaluate second\hadd{-}order corrections in XVH2 to motivate the need for appropriate approximation of \hadd{the} Green's function.

The second\hadd{-}order correction to \hadd{the} energy involves $2m$-dimensional integrals of the form, 
\begin{eqnarray}
I^{(2)} = \int \limits_{-\infty}^{+\infty}\int \limits_{-\infty}^{+\infty} e(\x,\x') P(\x,\x')
\dd\x \dd\x' \label{reform_E2}
\end{eqnarray}
with
\begin{eqnarray}
e(\x,\x') &=& \prod_{i=1}^m e^{-\omega_i(x_i^2+x'^{2}_i)}, \label{gaussian2} \\
\end{eqnarray}
where $\x = \{x_1,\dots,x_m\}$ is the $m$-dimensional set of normal coordinates, $\omega_i$ is the $i$th harmonic frequency which can be computed and is known a priori. The polynomial $P(\x,\x')$ given by
\begin{eqnarray}
P(\x,\x') &=& \Delta V(\x) \Delta V(\x') G(\x,\x'), \label{p2}
\end{eqnarray}
includes two functions. The first function, $\Delta V(\x)$, is the fluctuation potential given by
\begin{eqnarray}
\Delta V(\x) = V(\x) -V_{\mathrm{ref}} - \frac{1}{2} \sum_{i=1}^m \omega_i^2 x_i^2,
\end{eqnarray}
where $V(\x)$ is the $m$-dimensional PES and $V_{\mathrm{ref}}$ is its value at
the equilibrium geometry, which is the electronic energy at the equilibrium
geometry of the molecule. In this work, we are mainly concerned with the second function in $P(\x,\x')$ which is a real-space Green's function given by
\begin{eqnarray}
&& G(\x,\x') = \underset{(n_1,n_2,\dots,n_m) \neq (0,0,\dots,0)}{\sum_{n_1=0}^{n_{\mathrm{max}}-1}\cdots\sum_{n_m=0}^{n_{\mathrm{max}}-1}}\prod_{i=1}^m \frac{N^2_{n_i} h_{n_i}(\omega_i^{1/2}x_i)h_{n_i}(\omega_i^{1/2}x'_i)}
{-\sum_{i=1}^{m} n_i \omega_i} . \label{generalG}
\end{eqnarray}

Here, $N_{n_i}$ is a normalization coefficient, $h_{n_i}$ is the physicists' Hermite
polynomial of degree $n_i$ defined as $h_n(x)=(-1)^n e^{x^2} \frac{\dd^n}{\dd x^n}e^{-x^2}$. The highest quantum number for each quantum mode is $n_{max}\in \Nbb_0$. In general, higher the values of $n_{max}$, more accurate is the representation of Green's function for second order corrections in XVH2. In this study, we choose the same values of $n_{max}$ for $1\leq i\leq m$. We observe that the cost of storing coefficients $\frac{1}{-\sum_{i=1}^m n_i\omega_i}$ of Green's function in \eqeqref{generalG} is $(n_{max}^m-1)$, i.e. it scales exponentially with dimension $m$ which is a critical bottleneck in estimating \eqeqref{reform_E2}. This storage requirement can be easily seen by re-writing the above expression for $G(\x,\x')$ as 
\begin{equation*}
G(\x,\x') = \sum_{\nu\in\Delta} u_\nu g_\nu(\x,\x'),
\end{equation*}
where $\Delta=\{(n_1,\ldots,n_m)\,|\,n_i=1,\ldots,n_\mathrm{max};\, i=1,\ldots,m\} \setminus \{(0,\ldots,0)\}$, and
\begin{eqnarray*}
g_\nu(\x,\x') &=& \prod_{i=1}^m \phi^{(i)}_{n_i}(x_i,x_i')\\
\phi^{(i)}_{n_i}(x_i,x_i') &=&  N_{n_i} h_{n_i}(\omega_i^{1/2}x_i) N_{n_i} h_{n_i}(\omega_i^{1/2}x_i')\\
u_\nu &=& -\frac{1}{\sum_{i=1}^m n_i \omega_i}
\end{eqnarray*}
and noting the necessary storage for the $(n_{max}^m-1)$ coefficients $u_\nu$.

To estimate \eqeqref{reform_E2}, we search for low-rank approximations of the integrand factors, specifically

\begin{eqnarray}
\Delta V(\x) \approx \sum_{k=1}^{r_1} \prod_{i=1}^m \Delta V^{(i)}_k(x_i)
\label{lr_dv}
\end{eqnarray}
with a separation rank $r_1$ and
\begin{eqnarray}
G(\x,\x') \approx \sum_{k=1}^{r_2} \prod_{i=1}^m G^{(i)}_k(x_i,x'_i),
\label{lowrankG}
\end{eqnarray}
with a separation rank $r_2$. The functions $\Delta V^{(i)}_k(x_i)$ and $G^{(i)}_k(x_i,x'_i)$ are $k-\mathrm{th}$ univariate and bivariate functions in dimension $i$. Substituting \eqeqref{lr_dv} and \eqeqref{lowrankG} in \eqeqref{reform_E2}, the integral $I^{(2)}$ can be evaluated as a sum-of-products of two-dimensional integrals,
\begin{equation}
I^{(2)} \approx \sum_{k_1=1}^{r_1} \sum_{k_2=1}^{r_1} \sum_{k_3=1}^{r_2} \prod_{i=1}^m
\int_{-\infty}^{+\infty} \int_{-\infty}^{+\infty} e^{-\omega_i(x_i^2+x_i'^2)}
\Delta V^{(i)}_{k_1}(x_i) \Delta V^{(i)}_{k_2}(x'_i) \, G_{k_3}^{(i)}(x_i,x'_i) \,\dd x_i \,\dd x_i',
\label{sep_E2}
\end{equation}
The number of two dimensional integrals in \eqeqref{sep_E2} is $O(r_1^2r_2m)$ which can be evaluated using Gauss-Hermite quadrature. Since $p$ quadrature points can exactly evaluate integral of a polynomial of order $2p-1$, the computational cost of estimating \eqeqref{sep_E2} scales as $O(r_1^2r_2mp)$.

For accurate, efficient and scalable computation of  $I^{(2)}$ using separated integration with \eqeqref{sep_E2}, we require two conditions to be satisfied. Firstly, $\Delta V(\x)$ must be accurate in the form \eqeqref{lr_dv} with a small separation rank $r_1$. Secondly, separation rank $r_2$ in \eqeqref{lowrankG} must be small for sufficiently accurate approximation of Green's function. This requires not only accurate approximation of Green's function in a suitable low rank tensor format, but also efficient storage due to exponential increase in number of coefficients in $G(\x,\x')$ with $m$. In the following section, we detail our approach that satisfies these conditions. In this work, we propose a suitable strategy to store $G(\x,\x')$ such that its low rank approximation of the form \eqeqref{lowrankG} can be obtained with standard numerical schemes. 

\section{Tensor representation and low rank approximation of Green's function}
\label{sec:lowrankG}

The Green's function $G(\x,\x')$ is a coupled $2m$-dimensional function and, therefore, needs to be low-rank decomposed in the form \eqeqref{lowrankG} for estimating $I^{(2)}$. If one chooses as basis functions,
\begin{eqnarray}
\phi^{(i)}_{n_i}(x_i,x_i') =  N_{n_i} h_{n_i}(\omega_i^{1/2}x_i) N_{n_i} h_{n_i}(\omega_i^{1/2}x_i'),
\label{basis_G}
\end{eqnarray}
it is already formally decomposed as
\begin{eqnarray}
G(\x,\x') = \sum_{k=1}^{N_{max}} u_{k} \prod_{i=1}^m \phi^{(i)}_{n_i}(x_i,x_i'),
\label{multivariate_G}
\end{eqnarray}
where $k= k(n_1,\dots,n_m)$ is a counting index of the quantum numbers of modes 1 through $m$, corresponding to an ordering of multi-indices $(n_1,\dots,n_m)$,
and the expansion coefficient is known {\it a priori} as
\begin{eqnarray}
u_{k} = -\frac{1}{\sum_{i=1}^m n_i \omega_i}. \label{vk}
\end{eqnarray}

Here, $N_{max} = n_{max}^m-1$, where $n_{\mathrm{max}}$ is the highest quantum number of the harmonic-oscillator wave function included along each mode. Let $\Uc\in \otimes_{i=1}^m\Rbb^{n_{max}}$ denote the tensor of coefficients with components $\Uc_{n_1,\ldots,n_m} = \frac{1}{-\sum_{i=1}^m n_i \omega_i}$. Let us also represent $\Bc^s \in \otimes_{i=1}^m\Rbb^{n_{max}}$ as the tensor of basis functions evaluated at a sample realization $(\vx^s, \vx'^s)$ of $(\x, \x')$ such that $\Bc^s_{n_1,\ldots,n_m} = \prod_{i=1}^m \phi^{(i)}_{n_i}(\mathrm{x}^s_i,\mathrm{x}'^s_i)$. Evaluation of the Green's function $G(\vx^s,\vx'^s) \in \Rbb$ can then be represented as
\begin{eqnarray}
G(\vx^s,\vx'^s) = \langle \Uc,\Bc^s \rangle_{\setminus 0} = \underset{(n_1,n_2,\dots,n_m) \neq (0,0,\dots,0)}{\sum_{n_1=0}^{n_{\mathrm{max}}-1}\cdots\sum_{n_m=0}^{n_{\mathrm{max}}-1}}\prod_{i=1}^m \frac{N^2_{n_i} h_{n_i}(\omega_i^{1/2}\mathrm{x}_i^s)h_{n_i}(\omega_i^{1/2}\mathrm{x'}_i^{s})}
{-\sum_{i=1}^{m} n_i \omega_i},
\end{eqnarray}
where $\langle\cdot,\cdot\rangle_{\setminus 0}$ is the canonical inner product in $\otimes_{i=1}^m \Rbb^{n_{max}}$ with the exclusion of index corresponding to $n_i=0,1\leq i\leq m$. In this work, we treat this specific exclusion in Section \ref{sec:hierarchicalG}. Clearly, we need $O(n_{max}^m)$ computation for evaluation of $G(\vx^s,\vx'^s)$. If $n_{max}$ or $m$ is large, this cost is prohibitive for application of most methods, e.g. Monte Carlo, for estimation of $I^{(2)}$. Also, in this representation, given an \emph{a priori} choice of the basis tensor $\Bc$, we can identify $G(\x,\x')$ with the coefficient tensor $\Uc$. In the following we discuss low rank approximation of $\Uc$ in several tensor formats in order to approximate $G(\x,\x')$.

\begin{figure}[htb!]
\includegraphics[scale=0.6]{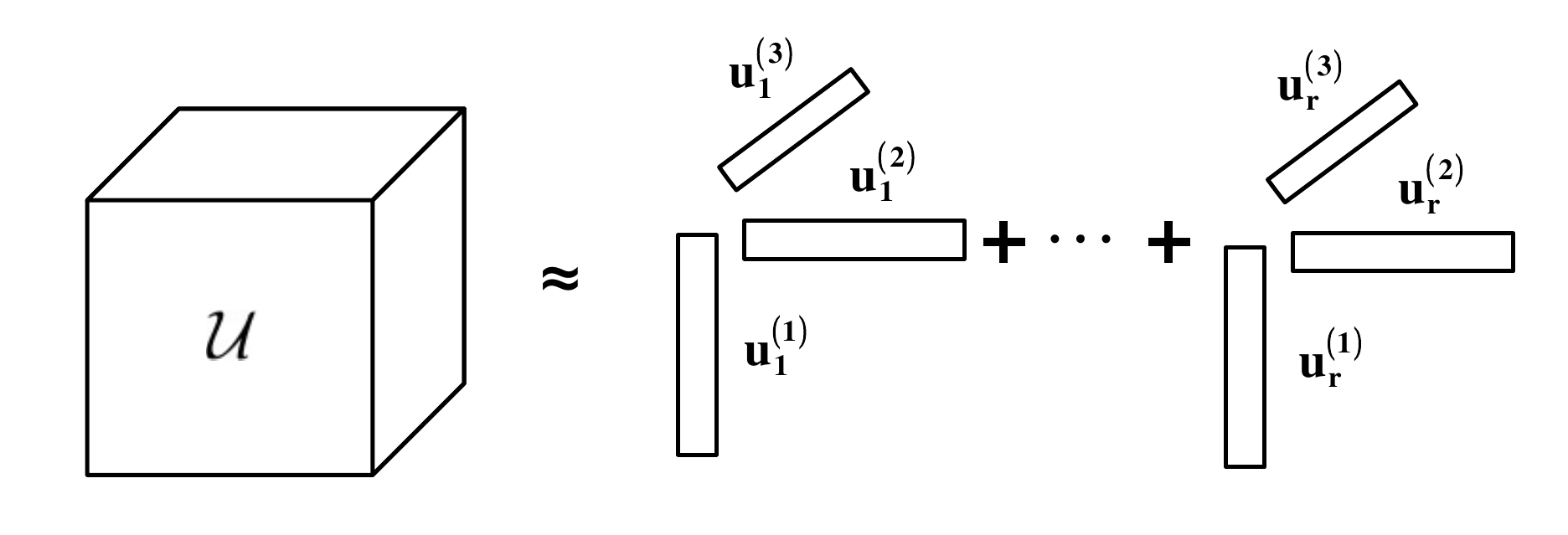}
\caption{Illustration of canonical tensor decomposition of $\Uc$}
\label{fig:can}       
\end{figure}

In our previous work \cite{RAI17,RAI19}, we approximated $G(\x,\x')$ by approximating $\Uc$ in the canonical polyadic tensor format
\begin{eqnarray}
\Uc \approx \Uc_{CP}=\sum_{k=1}^{r} \alpha_k (\otimes_{i=1}^m \u_k^{(i)}),
\end{eqnarray}
where $r$ is the separation rank and $\alpha_k$ is the normalization constant obtained by normalizing $\u_k^{(i)}, 1\leq i\leq m$. The corresponding functional representation of $G(\x,\x')$ is therefore given by
\begin{eqnarray}
G(\x,\x') \approx \sum_{k=1}^{r} \alpha_k \prod_{i=1}^m G^{(i)}_k(x_i,x'_i), \\ G^{(i)}_k(x_i,x'_i) = \langle\u_k^{(i)},\phib^{(i)} (x_i,x'_i)\rangle,
\end{eqnarray}
where $\phib^{(i)}(x_i,x'_i)$ is the vector of basis functions given by $(\phi^{(i)}_{n_1},\ldots,\phi^{(i)}_{n_{max}})^{T}$. The number of parameters in canonical polyadic tensor approximation of $G(\x,\x')$ is therefore given by $mrn_{max}$, which is linear in $m$. Figure \ref{fig:can} illustrates approximation of $\Uc$ in canonical tensor format.

\begin{figure}[htb!]
\includegraphics[scale=0.6]{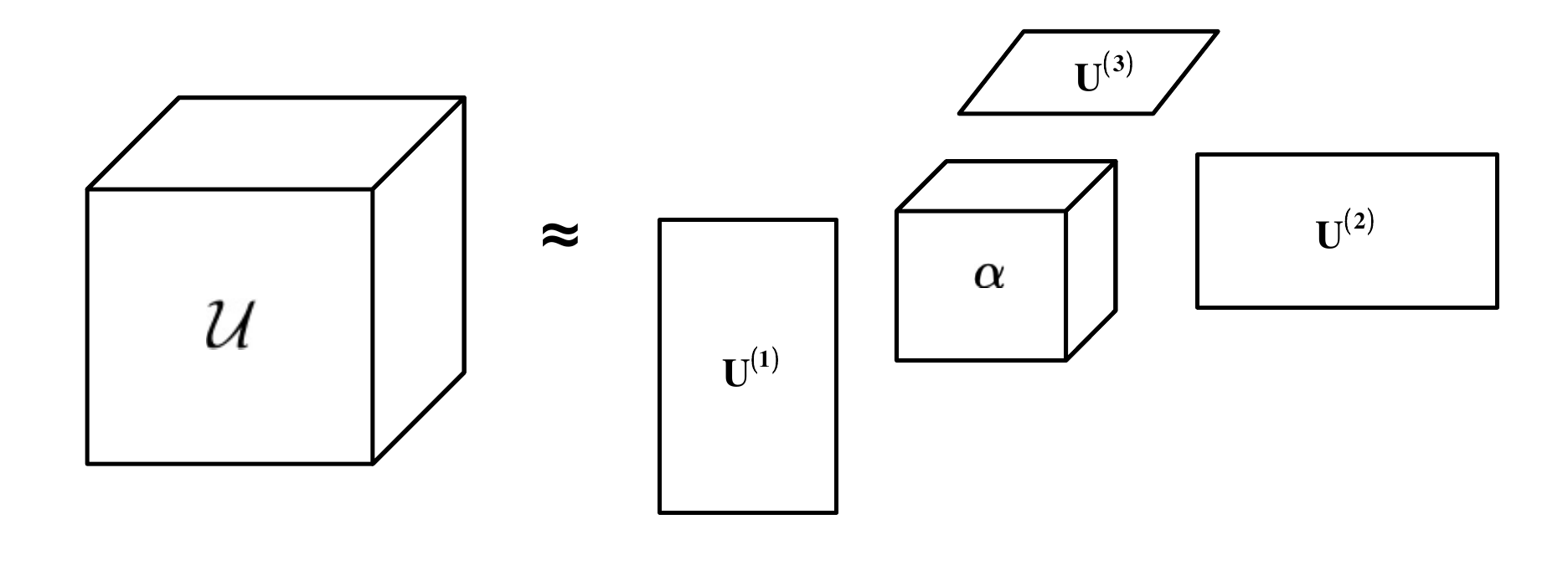}
\caption{Illustration of Tucker tensor decomposition of $\Uc$}
\label{fig:tuck}       
\end{figure}

Another format to approximate $\Uc$ is the Tucker tensor format which is represented as
\begin{eqnarray}
\Uc \approx \Uc_{T} = \sum_{k_1=1}^{r_1}\cdots\sum_{k_m=1}^{r_m} \alpha_{k_1,\ldots,k_m}\left(\otimes_{i=1}^m \u_{k_i}^{(i)}\right),
\end{eqnarray}
where $\alpha_{k_1,\ldots,k_m}$ form components of the core tensor $\alphab\in \Rbb^{r_1\times\cdots \times r_m}$   and $\u^{(i)}_{k_i} \in \Rbb^{n_{max}}, 1\leq k_i\leq r_i$ are columns of factor matrices $\U^{(i)}\in \Rbb^{n_{max}\times r_i}$ (see illustration below). Thus, as compared to canonical rank $r$ of $\Uc_{CP}$, which is a scalar, the multilinear rank of $\Uc_{T}$ is given by the tuple $(r_1,\ldots,r_m)$. An algorithm to approximate $\Uc$ in Tucker tensor format is called the higher order singular value decomposition (HOSVD) \cite{Lathauwer00}. This algorithm is based on the idea of minimal subspace. The minimal subspace, for a given mode $i$, is the minimum set of basis vectors that span the column space of $i^{th}$ mode unfolding of $\Uc$, where the $i^{th}$ mode unfolding is obtained by considering the $i^{th}$ mode as the first dimension of a matrix and collapsing $(1,\ldots,m)\setminus i$ as the other dimension. Practically, the minimal subspace is obtained using singular value decomposition (SVD) of the $i^{th}$ mode unfolding for $1\leq i\leq m$. Thus, for a given mode $i$, we first matricize $\Uc$ by reshaping as $\Uc \rightarrow \U_{i,{(1,\ldots,m)\setminus i}}$ and then perform SVD such that
\begin{eqnarray}
\U_{i,{(1,\ldots,m)\setminus i}} \approx \sum_{k_i=1}^{r_i} \beta_i \u_{k_i}^{(i)}\otimes \v_{k_i}^{(1,\ldots,m)\setminus i},
\end{eqnarray}
where the left singular vectors $\u^{(i)}_{k_i}$ form the column of factor matrix $\U^{(i)}$, $\v_{k_i}^{(1,\ldots,m)\setminus i}$ are right singular vectors and $r_i$ is the corresponding component of Tucker rank. The core tensor $\alphab$ is calculated by projecting $\Uc$ on each subspace separately. The number of parameters in the Tucker tensor approximation of $\Uc$ is the sum of size of the core tensor $\prod_{i=1}^m r_i$ and size of the factor matrices $\sum_{i=1}^m r_in_{max}$. The size of the core tensor is thus exponential in $m$ which limits the Tucker decomposition of $\Uc$ to small molecules. Figure \ref{fig:tuck} illustrates the approximation of $\Uc$ in the Tucker tensor format.

Approximation of $\Uc$ in canonical or Tucker formats first requires the storage of $\Uc$, the size of which increases exponentially with $m$. Thus, for any $n_{max}>1$, the storage cost of $\Uc$ will become infeasible for large enough values of $m$. To overcome this limitation, we propose to store and decompose $\Uc$ in tree based tensor formats, called Hierarchical Tucker format, that is based on recursive application of Tucker decomposition to a hierarchy of partition of dimensions.

\begin{figure}[h!]
\centering
\includegraphics[scale=.50]{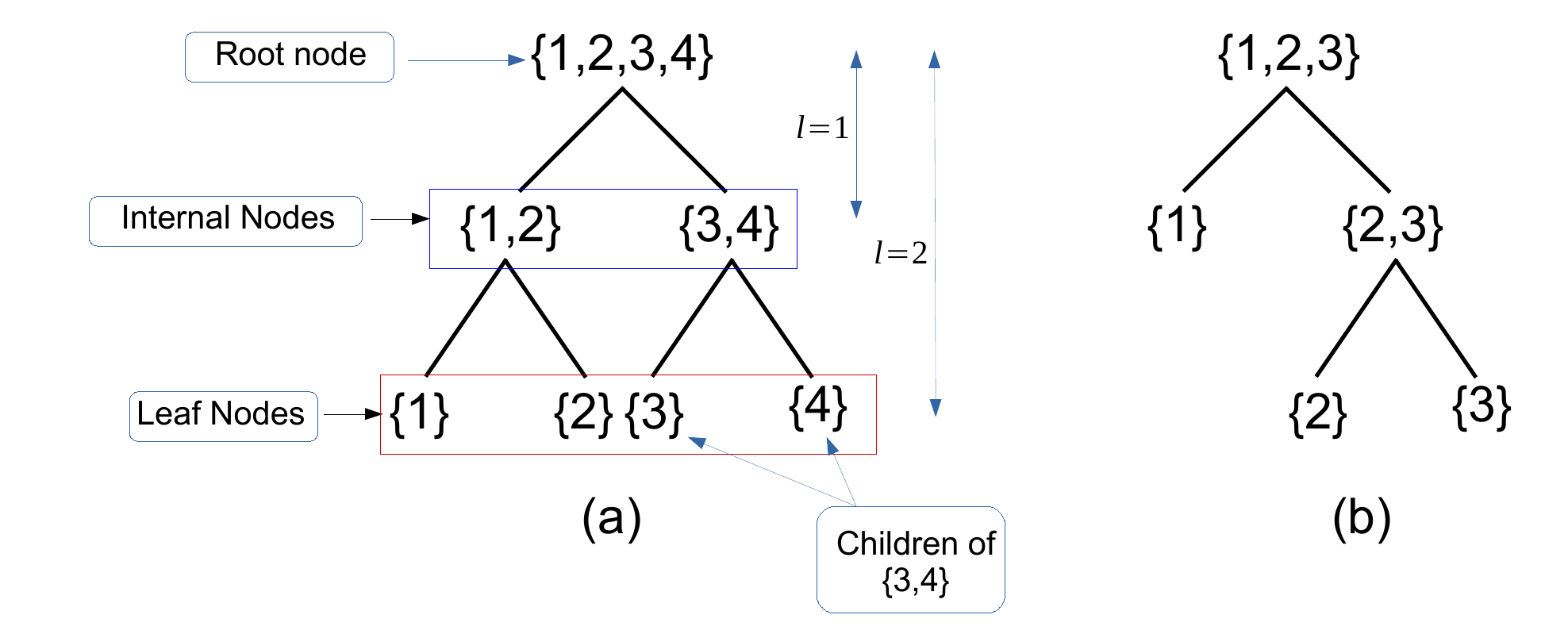}
\caption{(a) An example dimension tree for $m=4$. (b) Dimension tree for $m=3$ which also corresponds to the tensor train representation}
\label{dim_tree}
\end{figure}

Tree based formats are based on a more general notion of rank for a group of dimensions associated with a dimension  tree. Let $M=\{1,\ldots,m\}$ and $T$ be a dimension partition tree on $M$, such that every vertex $t\in T$ are non empty subsets of $M$ (see Figure \ref{dim_tree} for illustration). Let us denote $L(T)$ as the leaves of $T$ and $I(T)=T\setminus L(T)$, so that $L(T)=\{\{k\}:k\in M\}$. Let $S(t)$ denote children of $t\in T$. A node is said to be at a level $l$ if it has a distance of exactly $l$ to the root $M$.
Hierarchical Tucker tensor format is associated with a binary tree $T$, such that for all $t\in I(t)$, $\#S(t)=2$ i.e. all nodes other than leaf nodes have two children.

\begin{figure}[h!]
\centering
\includegraphics[scale=.75]{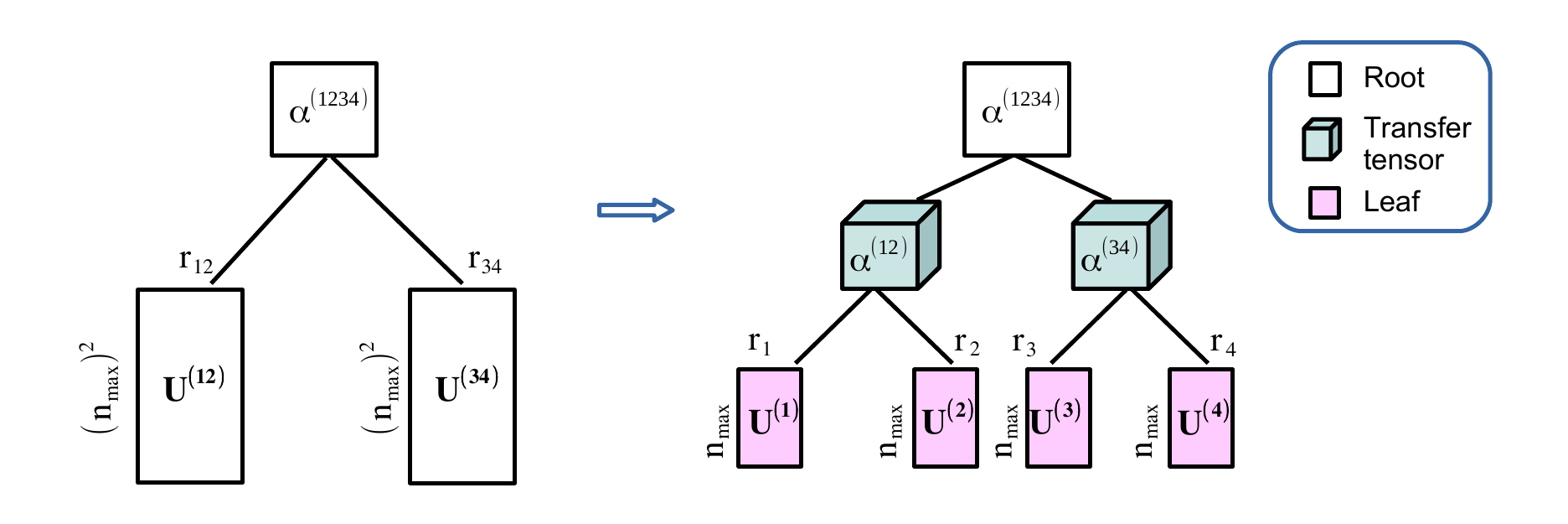}
\caption{Illustration of steps in construction of Hierarchical Tucker tensor Format for dimension $m=4$ corresponding to dimension tree in Figure \ref{dim_tree}(b)}
\label{fig:hierarchical}
\end{figure}

In this tensor format, we have rank associated with each vertex $t$. Therefore, the Hierarchical tensor rank of $\Uc$ is a tuple $(\text{rank}_t(\Uc))_{t\in T}\in \Nbb^{\#T}$ such that $\text{rank}_t(\Uc)=\text{\#columns}(\mathbf{U}^{(t)})$, where $\mathbf{U}^{(t)}$ is the minimal subspace for dimensions associated with vertex $t$. Let us denote by $(\u_k^{(t)})_{1\leq k\leq r_t}$ as columns of $\U^{(t)}$. For $t\in I(t),$ with $S(t)=\{t_1,t_2\},$ we can write
\begin{equation}
\u^{(t)}_{k}=\underset{\substack{1\leq l\leq r_{t_1} \\ 1\leq r \leq r_{t_2}}}\sum \alpha^{(t)}_{klr}\u^{(t_1)}_{l}\otimes \u^{(t_2)}_{r}.
\end{equation}
for $1\leq k\leq r_t$. The tensor $\alphab^{(t)}\in \Rbb^{r_t\times r_{t_1}\times r_{t_2}}$ are called the \textit{transfer tensors} with components $\alpha^{(t)}_{klr}$. With $r_M=1$, approximation of $\Uc$ in Hierarchical tensor format is represented as
\begin{equation}
\Uc=\sum_{l=1}^{r_{M_1}}\sum_{r=1}^{r_{M_2}}\alpha_{lr}^{(M)}\u_l^{(M_1)}\otimes \u^{(M_2)}_r,
\end{equation}
where $S(M)=\{M_1,M_2\}$. Thus, the tensor $\Uc$ is completely determined by the transfer tensors $(\alphab^{(t)})_{t\in I(T)}$ and the vectors $(\u^{(i)}_k)_{k\in L(T),1\leq k\leq r_k}$. Figure \ref{fig:hierarchical} illustrates Hierarchical tensor format for a tensor corresponding to dimension partition tree in Figure \ref{dim_tree} (a). Note that a given tensor can be approximated in several Hierarchical Tucker tensor each associated with a different dimension tree.

The complexity of $\Uc$ in Hierarchical format includes the storage cost of parameters in leaf nodes and internal nodes, which includes the root node and transfer tensors. We have $\sum_{i=1}^m n_{max} r_i$ parameters in the leaf nodes and $\sum_{t\in I(T)}r_t r_{t_1} r_{t_2}$ in internal nodes. There are therefore $\sum_{i=1}^m r_mn_{max}+\sum_{t\in I(T)}r_t r_{t_1} r_{t_2}$ scalar parameters with storage cost linear in $m$.

In the following section, we present a method to efficiently store and approximate $\Uc$ in Hierarchical format.

\section{Storage and Approximation of Green's function in Hierarchical Format}
\label{sec:hierarchicalG}

Let us consider a tensor $\Uc^{*}$ that contains element-wise reciprocal of components of $\Uc$ i.e.
\begin{eqnarray}
\Uc^{*}_{n_1,\ldots, n_m} = \frac{1}{\Uc_{n_1,\ldots, n_m}}.
\end{eqnarray}

We can represent $\Uc^{*}$ exactly in Hierarchical format for any dimension tree (see section 8 and 9 in \cite{kressner2012}) such that the leaf nodes $\U^{(i)} \in \Rbb^{n_{max}\times 2}$ are given by
\begin{eqnarray}
\U^{(i)} =
\begin{bmatrix}
1 & n_1\omega_i \\
\vdots & \vdots \\
1 & n_{max}\omega_i
\end{bmatrix},
\end{eqnarray}
the transfer tensors $\alphab^{(t)}\in \Rbb^{2\times 2\times 2}$ for $t\in I(T)\setminus t_{root}$ are given by
\begin{eqnarray}
\alphab^{(t)}_{:,:,1} =
\begin{bmatrix}
1 & 0 \\
0 & 0
\end{bmatrix},\;\;
\alphab^{(t)}_{:,:,2} =
\begin{bmatrix}
0 & 1 \\
1 & 0
\end{bmatrix},
\end{eqnarray}
and the root node $\alphab_{t_{root}} \in \Rbb^{2\times 2}$ is given by
\begin{eqnarray}
\alphab_{t_{root}} =
\begin{bmatrix}
0 & 1 \\
1 & 0
\end{bmatrix}.
\end{eqnarray}
We denote $\Uc^{(*)}$ stored in Hierarchical tensor format as $\Uc^{(*)}_H$. Let us illustrate the storage of $\Uc^{*} \in \Rbb^{2\times 2\times 2}$ as $\Uc^{(*)}_H$ with an example. We consider a case with $m=3$, $n_{max} = 2$ and the dimension tree as shown in Figure \ref{dim_tree} (b). Note that this dimension tree also corresponds to a particular case of Hierarchical format called the tensor train format \cite{Oseledets:2011}.
%

\begin{figure}[htb!]
\begin{center}
\includegraphics[scale=1.2]{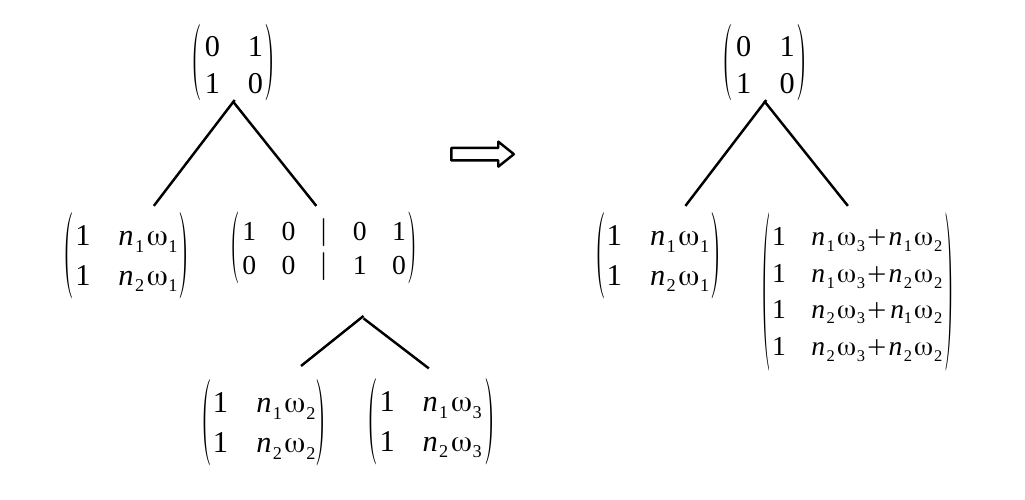}
\end{center}
\caption{Illustration of storage of $\Uc^{(*)}$ with $m=3, n_{max} = 2$ in Hierarchical Tucker tensor format and its unfolding from level two to level one.}
\label{fig:store_level_1}
\end{figure}

Figure \ref{fig:store_level_1} shows unfolding of $\Uc_H^{(*)}$ from level two to level one. We have

\begin{eqnarray}
\u^{(23)}_{1} = \sum_{l=1}^{2}\sum_{r=1}^{2} \alphab^{(23)}_{:,:,1} \u^{(2)}_l\otimes \u^{(3)}_r =
\begin{bmatrix}
1 & 1 \\
1 & 1
\end{bmatrix}.
\end{eqnarray}
Similarly, we have
\begin{eqnarray}
\u^{(23)}_{2} = \sum_{l=1}^{2}\sum_{r=1}^{2} \alphab^{(23)}_{:,:,2} \u^{(2)}_l\otimes \u^{(3)}_r =
\begin{bmatrix}
n_1\omega_3 + n_1\omega_2 & n_2\omega_3 + n_1\omega_2\\
n_1\omega_3 + n_2\omega_2 & n_2\omega_3 + n_2\omega_2
\end{bmatrix}.
\end{eqnarray}

\begin{figure}[htb!]
\begin{center}
\includegraphics[scale=0.4]{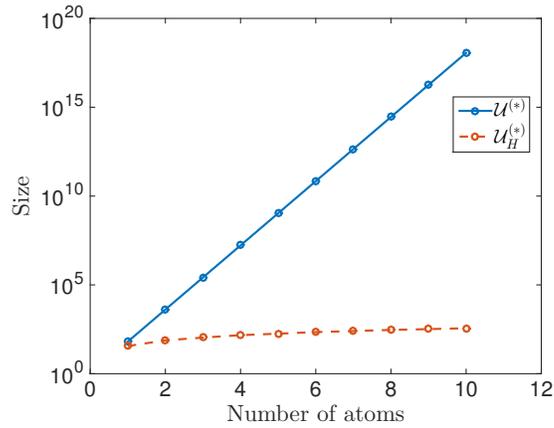}
\end{center}
\caption{Scaling of the storage cost of $\Uc^{(*)}$ and $\Uc^{(*)}_H$ with $n_{max} = 4$  v/s number of atoms in a molecule.}
\label{fig:scaling}
\end{figure}

Both $\u_1^{(23)}$ and $\u_2^{(23)}$ can be reshaped in $\Rbb^{n_{max}^2\times 1}$ to obtain columns of $\U^{(23)}$. A similar unfolding can be done from level 1 to level 0 such that

\begin{eqnarray*}
&\Uc^{(*)} = \sum_{l=1}^{2}\sum_{r=1}^{2} \alphab^{(123)} \u^{(1)}_l\otimes \u^{(23)}_r \\
&= \begin{bmatrix}
n_1\omega_3 + n_1\omega_2  + n_1\omega_1 &  n_1\omega_3+n_2\omega_2+n_1\omega_1 & n_2\omega_3 + n_1\omega_2 + n_1\omega_1 & n_2\omega_3+n_2\omega_2+n_1\omega_1\\
n_1\omega_3 + n_1 \omega_2 +n_2\omega_1  & n_1\omega_3 + n_2\omega_2 + n_2\omega_1 &  n_2\omega_3+n_1\omega_2+n_2\omega_1 &n_2\omega_3 + n_2\omega_2 + n_2\omega_1
\end{bmatrix},
\end{eqnarray*}

which can then be folded as $\Uc^{(*)} \in \Rbb^{2\times 2\times 2}$. This illustration shows that $\Uc^{(*)}$ can be stored efficiently in Hierarchical tensor format. Clearly, the benefit of storing $\Uc^{(*)}$ as $\Uc^{(*)}_H$ is significant for higher values of $m$ and $n_{max}$. Figure \ref{fig:scaling} compares the scaling of storage cost of $\Uc^{(*)}$ and $\Uc^{(*)}_H$ for $n_{max} = 4$ (for a dimension tree with two levels, level two being leaf nodes and nodes at level one being those of cardinality two) with respect to number of atoms in the molecule. We find that, for a molecule with only 10 atoms, the storage of $\Uc^{(*)}$ becomes practically infeasible whereas $\Uc^{(*)}_H$ requires storage of only 364 real values based on storage calculation of Hierarchical Tucker Tensor formats indicated in Section \ref{sec:lowrankG}.

Next, starting with $\Uc^{(*)}_H$, we wish to obtain an approximation of $\Uc$ in Hierarchical tensor format, henceforth denoted as $\Uc_H$. This involves application of an algorithm that results in element wise reciprocal of $\Uc^{(*)}$ while maintaining its representation as $\Uc^{(*)}_H$. In the context of tree based tensor formats, such an algorithm based on Newton Schulz iterations has already been proposed. We refer the reader to \cite{kressner2012} and \cite{TTToolbox} for this algorithm and its implementation details. We now deal with the issue of exclusion of coefficient corresponding to $n_i=0, 1\leq i\leq m$ in $\Uc$. Since $\Uc$ is never explicitly created in this approach, the first coefficient of $\Uc$ corresponding to $n_i=0, 1\leq i\leq m$ cannot be explicitly assigned to 0. However, once we get $\Uc_H$, we can get the coefficient estimated by Newton Schulz corresponding to this entry, find its representation in Hierarchical Tucker format with the same dimension tree as that of $\Uc_H$ and subtract it explicitly from $\Uc_H$.

\begin{figure}[h!]
\centering
\includegraphics[scale=.50]{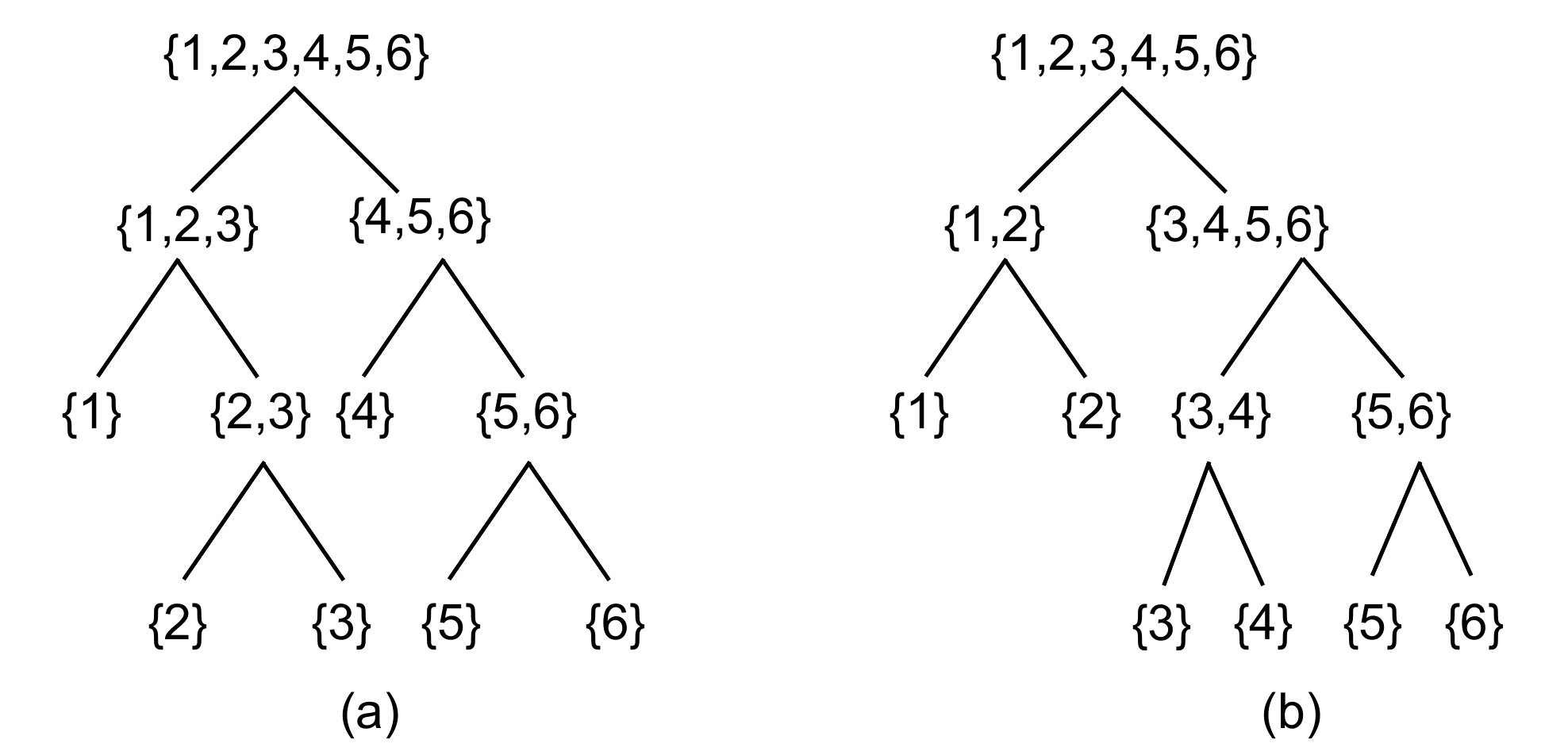}
\caption{Dimension Tree 1 (left) and Tree 2 (right) for approximation of $\Uc_H$.}
\label{tree_types}
\end{figure}

To illustrate the approximation of $\Uc_H$ from $\Uc^{(*)}_H$, we consider Green's function for four different molecules: water $(m=3)$, formaldehyde ($m=6$), methane ($m=9$) and ethylene ($m=12$). For water and formaldehyde, we consider $n_{max} = 4,8$. For methane and ethylene, due to limitations associated with storage of the full tensor $\Uc$ in order to estimate approximation error of $\Uc_H$, we only consider $n_{max}=4$. For all molecules, we also consider two different dimension trees (see Figure \ref{tree_types}) to illustrate that the accuracy of approximation of $\Uc_H$ is not sensitive to structure of the tree. For both trees, at a given node, the dimensions on the left subtree are smaller than the ones on the right. In case the cardinality of a given node is odd, we split this node such that the cardinality of its left child is one smaller than the cardinality its right child. The second tree differs from the first in the sense that is root node has two leaf nodes corresponding to the leading two dimensions. We illustrate these two types of dimension trees for a molecule with four atoms (i.e, $m=6$) in Figure \ref{tree_types}. Table \ref{tab:rel_u_uH} shows relative error $\frac{\Vert\Uc-\tilde{\Uc}\Vert}{\Vert\Uc\Vert}$ in approximation of $\Uc$ and its approximation $\tilde{\Uc}$ obtained by unfolding $\Uc_H$. Here, the norm $\Vert\cdot\Vert$ is the canonical norm in $\otimes_{i=1}^{m}\Rbb^{n_{max}}$, which is calculated in practice by taking the square root of the sum of squared entries of the tensor. 

We clearly find that accurate approximations of $\Uc$ of the order $10^{-9}$ to $10^{-11}$ are obtained for all four molecules. Also, the choice of tree has little influence on the accuracy of approximation. In the following, we present a modification of this approximation for application in separated integration to determine second order anharmonic energy correction of molecules.

\begin{table}[htb!]
\begin{center}
\caption{Relative error $(\times 10^{9})$ in approximation of $\Uc$ from $\Uc_H$ for water, formaldehyde, methane and ethylene. The error is reported for approximation corresponding to two different trees in Figure \ref{tree_types}. For $\mathrm{CH_4}$ and $\mathrm{C_2H_4}$, $n_{max}=8$ is not considered due to storage issues associated with $n_{max}^m$ coefficients.}
\label{tab:water}
\begin{tabular}{p{1.5cm}c c c c c c}
\hline\noalign{\smallskip}
 & \multicolumn{2}{c}{$\mathrm{H_2O}$}& \multicolumn{2}{c}{$\mathrm{CH_2O}$} & $\mathrm{CH_4}$ & $\mathrm{C_2H_4}$\\
\noalign{\smallskip}\hline\noalign{\smallskip}
$n_{max}$& 4 & 8 & 4 & 8 & 4 & 4\\
Tree 1 & 0.023 & 0.021 & 0.047 & 9.4 & 0.11 & 3.82\\
Tree 2 & 0.003 & 0.012 & 0.008 & 7.74 & 1.08 & 6.16\\
\noalign{\smallskip}\hline\noalign{\smallskip}
\end{tabular}
\label{tab:rel_u_uH}
\end{center}
\end{table}

\section{Second order anharmonic corrections with Green's function}
\label{sec:anharmonicG}

In this section, we first present results related to approximation of Green's function in canonical tensor format followed by  an illustration of accurate estimation of $I^{(2)}$. We show results for molecules considered in the previous section.

In order to estimate $I^{(2)}$ using \eqeqref{sep_E2}, we need to convert $G(\x,\x')$ represented in Hierarchical tensor format with $\Uc_H$ into canonical polyadic tensor format represented by $\Uc_{CP}$. Conversion between tensor formats is a standard operation in tensor methods and we refer the reader to \cite{Hackbusch:2012} for details. Let us define the approximation error $\epsilon$ such that

\begin{eqnarray}
\epsilon = \frac{\Vert \G - \tilde{\G}\Vert_2}{\Vert \G \Vert_2},
\end{eqnarray}
where $\G$ and $\tilde{\G}$ are vectors of evaluations of $G(\x,\x')$ and its canonical polyadic approximation respectively at $10,000$ uniformly distributed random samples of $\x,\x'$ over the same range as that over which $\Delta V(\x)$ is evaluated for estimating $I^{(2)}$.

\begin{figure}[htb!]
\centering
\includegraphics[scale=.37]{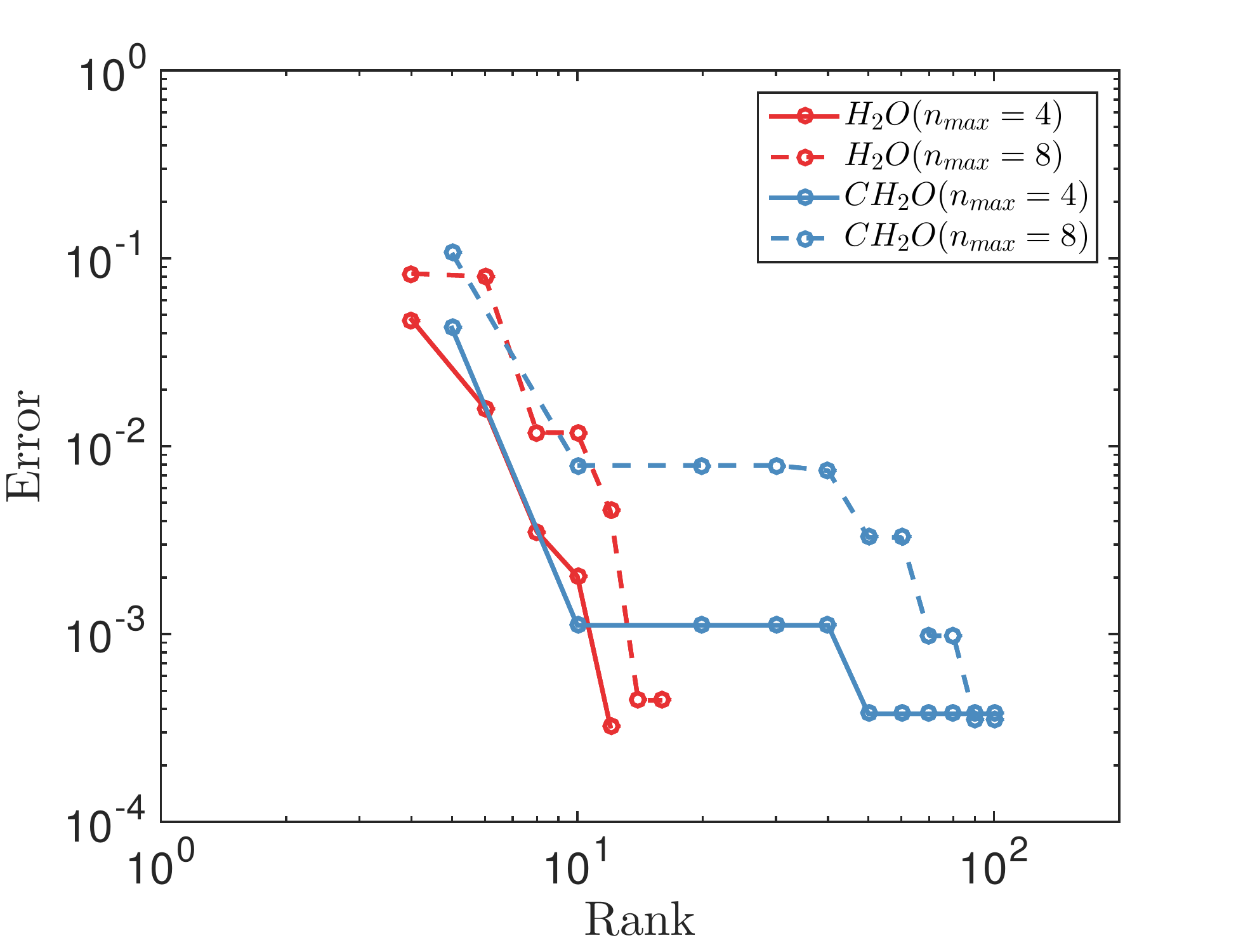}
\caption{Approximation error $\epsilon$ of Green's function for water ($m=3$) and formaldehyde ($m=6$) v/s canonical rank $r$ with $n_{max} = 4,8$. The error is estimated using $10,000$ independent evaluations of Green's function and its approximation in canonical polyadic tensor format.}
\label{fig:greenCanonical_w_f}
\end{figure}

Figure \ref{fig:greenCanonical_w_f} illustrates error in approximation of $G(\x,\x')$ in canonical tensor format obtained from $\Uc_H$ for water and formaldehyde. In \cite{RAI19}, we determined empirically that $\epsilon \approx 1.0\times 10^{-2}$ in approximation of integrand factors is sufficient for accurate estimates of second order corrections. In case of water, we get a sufficiently accurate approximation of $G(\x,\x')$ with a separation rank $r \approx 20$ for $n_{max} = 4,8$. Similarly for formaldehyde, we get $\epsilon \approx 1.0\times 10^{-2}$ for $r \approx 20$. In each case, as we increase the canonical rank, we get better approximation of $G(\x,\x')$. Also, for both molecules, the canonical rank $r$ required to achieve similar accuracy increases as the size of $\Uc$ increases with higher values of $n_{max}$.

\begin{figure}[htb!]
\centering
\includegraphics[scale=.37]{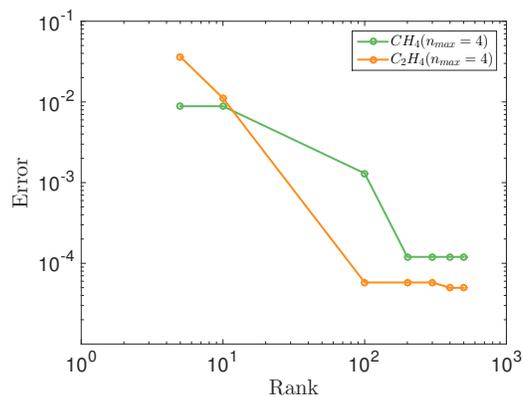}
\caption{Approximation error $\epsilon$ of Green's function for methane ($m=9$) and ethylene ($m=12$) v/s canonical rank $r$ with $n_{max}=4$. The error is estimated using $10,000$ independent evaluations of Green's function and its approximation in canonical polyadic tensor format.}
\label{fig:greenCanonical_m_e}
\end{figure}

In Figure \ref{fig:greenCanonical_m_e}, we find that for methane and ethylene, we get $\epsilon < 1.0\times 10^{-2}$ for a separation rank of $r=100$. For both these molecules, we illustrate approximation of $G(\x,\x')$ for $n_{max} = 4$ only due to limitations associated with storage and evaluation of exact value of $G(\x,\x')$ for higher values of $n_{max}$. We note that, for all molecules, we obtain small values of separation rank $r$ as compared to $\Oc(n_{max}^m)$ terms in $G(\x,\x')$.

Although $\epsilon \approx 1.0\times 10^{-2}$ is empirically determined to be sufficiently accurate for estimating $I^{(2)}$, we choose rank $r$ corresponding to a more conservative value of $\epsilon \approx 1.0\times 10^{-3}$.

\begin{figure}[htb!]
\includegraphics[scale=.37]{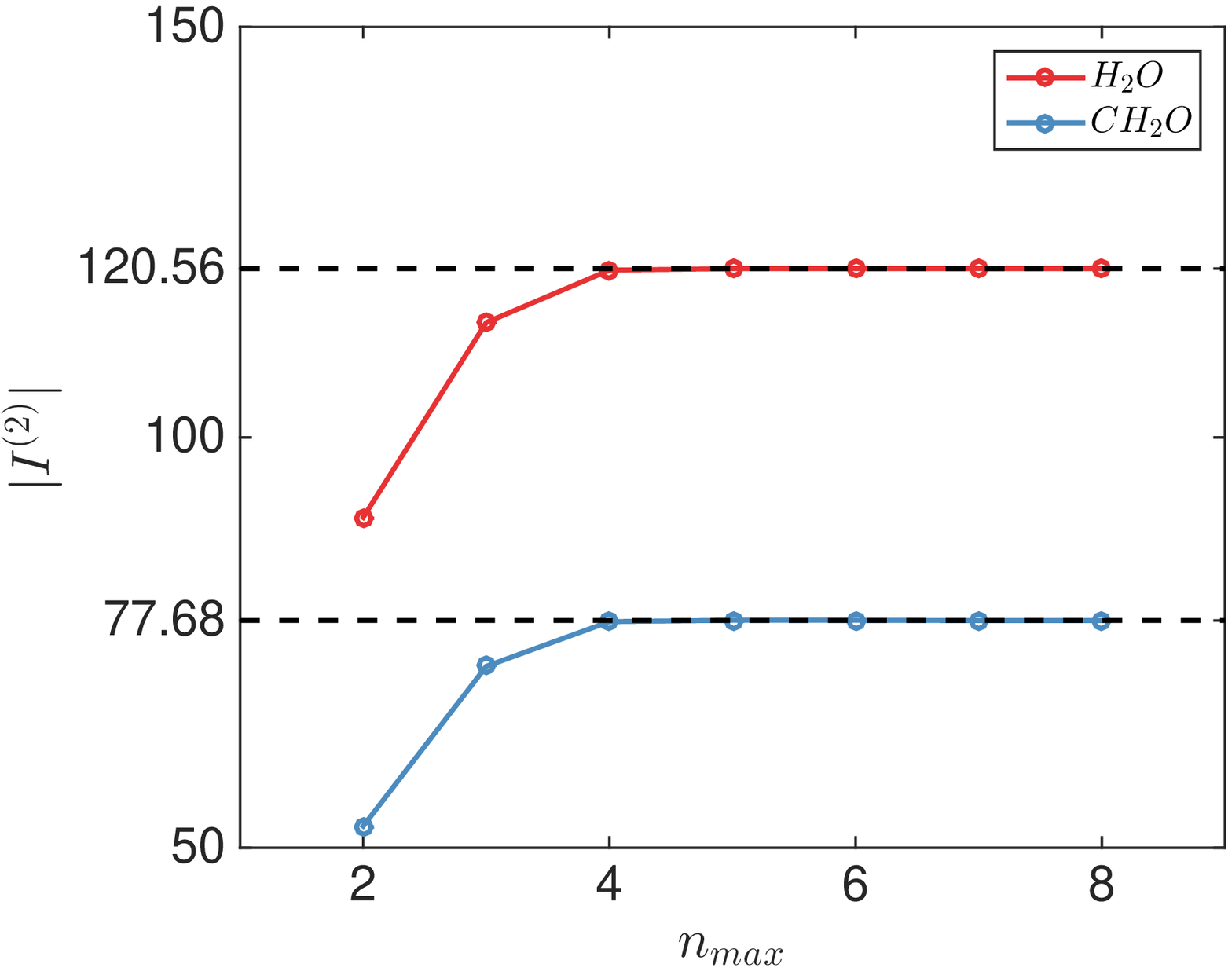}
\includegraphics[scale=.37]{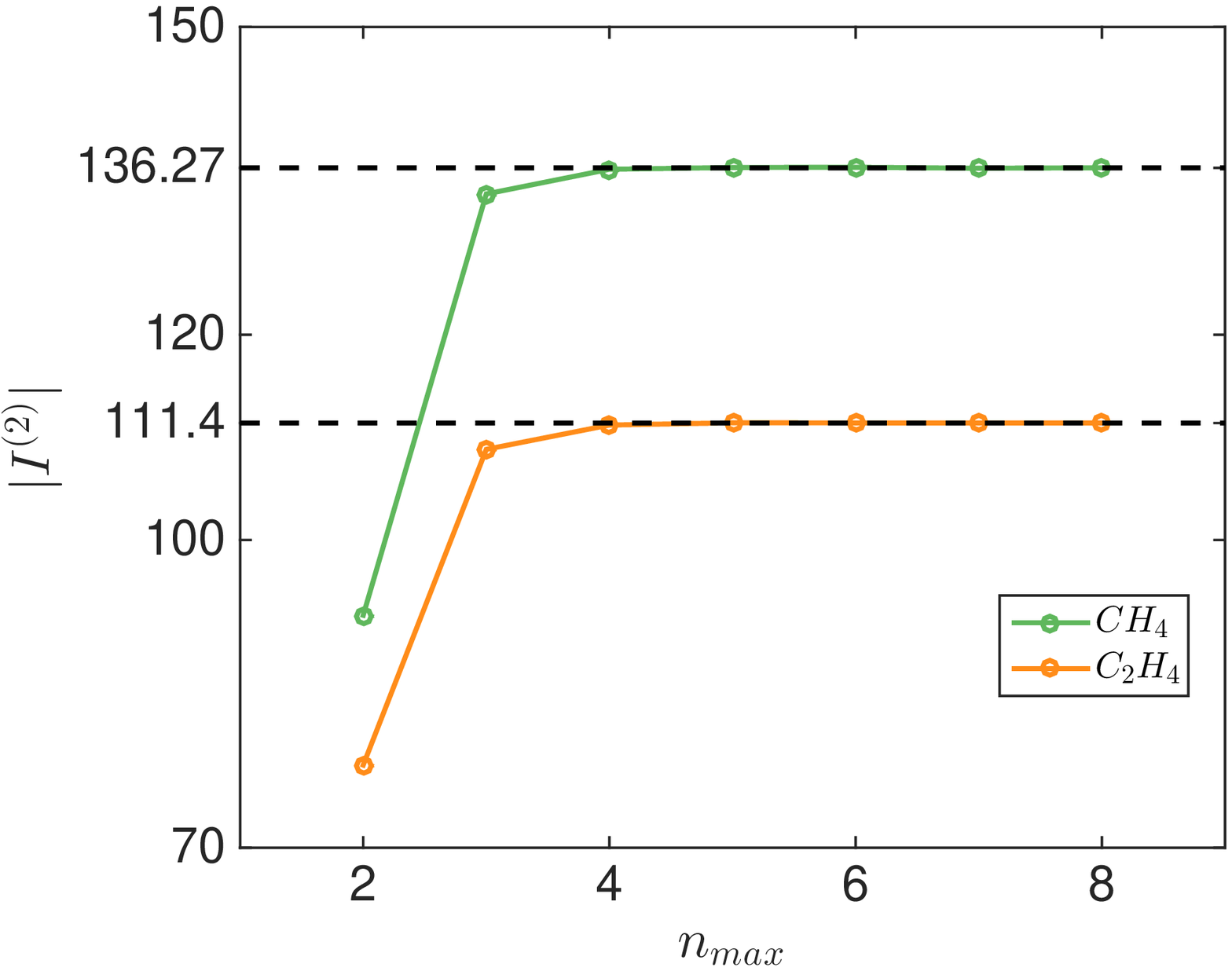}
\caption{Convergence of second order energy correction for water, formaldehyde (left), methane and ethylene (right) using low rank approximation of Green's function with quantum number $2\leq n_{max}\leq 8$. A fixed low rank representation of potential energy surface obtained with SCT-XVH2 \cite{RAI19} has been used for all molecules to illustrate dependence in accuracy on $n_{max}$.}
\label{fig:anharmonic}
\end{figure}

\begin{figure}[htb!]
\centering
\includegraphics[scale=.37]{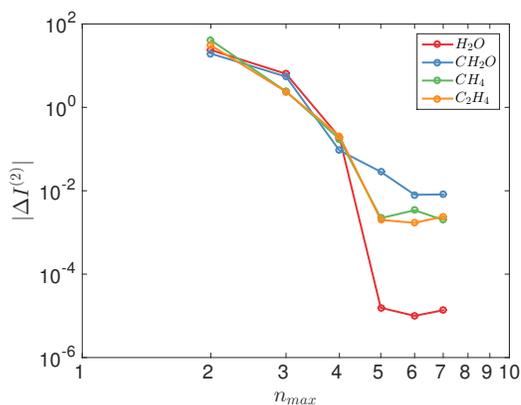}
\caption{Convergence plot of $|\Delta I^{(2)}| =  |I^{(2)}_{n_\mathrm{max}}-I^{(2)}_{n_\mathrm{max}+1}|$ vs $n_{max}$}
\label{fig:self_convergence}
\end{figure}

We show the convergence of $I^{(2)}$ with the approximation of $G(\x,\x')$ obtained with $2\leq n_{max}\leq 8$ in Figure \ref{fig:anharmonic}. Note that convergence $I^{(2)}$ does no depend only on $n_{max}$ but also on the approximation accuracy of potential energy surface $\Delta V(\x)$. Convergence plots in \ref{fig:anharmonic} have been obtained by fixing low rank approximations of $\Delta V(\x)$ in \eqeqref{lr_dv} obtained with SCT-XVH2 method proposed in our earlier work \cite{RAI19}. Figure \ref{fig:self_convergence} shows self convergence of $I^{(2)}$ with $n_{max}$ for all four molecules. We find that, for all four molecules, we get a convergence of the order at least $10^{-2}$, thus indicating monotonic increase in the accuracy in the approximation of $I^{(2)}$. We find that, for all molecules \hadd{considered here}, the absolute value of second order energy correction |$I^{(2)}$| estimated using \eqeqref{sep_E2} ceases to change appreciably for $n_{max}\ge 4$. Thus it is observed that the smallest value of $n_{max}$ required for estimation of second order energy corrections is independent of dimension $m$, although a definitive conclusion can only be drawn after a more comprehensive study involving molecules of different types and size. This result has two implications. Firstly, the size of \hadd{the} Green's function tensor to be approximated in low rank format will be smaller as compared to the one considered with higher values of $n_{max}$, a result that is especially significant for estimating corrections for bigger molecules. Secondly, with a fixed value of $n_{max} = 4$, the number of quadrature points and hence the computation cost required for numerical estimation of separated integrals in \eqeqref{sep_E2} will depend only on the degree of polynomial approximation of the PES.

\begin{figure}[htb!]
\centering
\includegraphics[scale=.45]{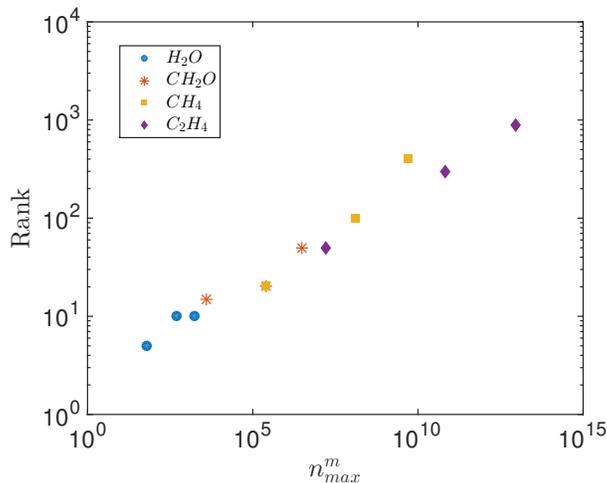}
\caption{Comparison of reduction in rank of Green's function from $n_{max}^m$ to canonical rank $r$ for $n_{max} \in \{4,8,12\}.$}
\label{fig:final_scaling}
\end{figure}

Finally, in Figure \ref{fig:final_scaling} we emphasize the advantage of proposed method by plotting reduction in rank of the Green's function in order to achieve $I^{(2)}  \pm 1.0\times 10^{-1}$ (where the value of $I^{(2)}$ for each molecule is shown in Figure \ref{fig:anharmonic} with dotted lines). We consider three values of $n_{max}\in\{4,8,12\}$ for each molecule. Consistent with our observation above, as the size of coefficient tensor increase with increase in values of $n_{max}$, we get higher values of canonical rank $r$ for each molecule. For the maximum value of $n_{max}^m = 10^{12}$ considered in Figure \ref{fig:final_scaling}, one can estimate accurate values of $I^{(2)}$ for molecules with up to 7 atoms (i.e. $m=21$) provided $I^{(2)}$ remains the same for $n_{max}\ge 4$ as is the case with molecules considered in this study. Note that, this method enables estimation of $I^{(2)}$ for molecule of any size provided accurate polynomial representation of PES can be obtained. In such cases, Figure \ref{fig:final_scaling} can also be extended to molecules bigger than those considered in this study.  
\section{Conclusion}
\label{sec:conclusion}
We presented a scalable method to efficiently store and compress the Green's function in the Hierarchical Tucker tensor format that overcomes a critical bottleneck in estimation of anharmonic corrections using XVH2. In particular, we showed that the complexity of Green's function in Hierarchical format is linear in dimension due to which it can be approximated for molecules of any size. We also illustrated that approximation of Green's function obtained using this approach can lead to reduction in separation rank by orders of magnitude and consequent reduction in computation cost for estimating anharmonic corrections for molecules. Finally, application of this method on molecules considered in this study suggests that a low rank approximation of Green's function with a small value of quantum number leads to an accurate approximation of second order energy corrections.

The proposed method can be enhanced along two directions. First, since the method depends on the availability of PES for estimating $I^{(2)}$, one can also approximate the PES of bigger molecules in Hierarchical Tensor format provided enough PES samples are available for approximation. Second, the present approach requires conversion of Green's function representation in Hierarchical Tensor format to canonical form for the application of quadrature integration rule. This intermediate step can avoided if the separated integration is performed directly in Hierarchical format.    

We used two open source software packages to implement and test methods proposed in this work. For approximation of $\Uc$ as $\Uc_{CP}$ and $\Uc_{T}$ in Section~\ref{sec:lowrankG}, we refer to Tensor Toolbox \cite{TTB_Software}. For operations related to storage and approximation of $\Uc_{H}$ in Section~\ref{sec:hierarchicalG} and Section~\ref{sec:anharmonicG}, we refer to htucker \cite{kressner2012}. 

\section{Acknowledgement}
Support for this work was provided through the Scientific
Discovery through Advanced Computing (SciDAC) program funded by the U.S.
Department of Energy, Office of Science, Advanced Scientific Computing Research
and Basic Energy Sciences under Award No.  DE-FG02-12ER46875. Sandia National
Laboratories is a multimission laboratory operated by National Technology and
Engineering Solutions of Sandia LLC, a wholly owned subsidiary of Honeywell
International Inc., for the U.S. Department of Energy's National Nuclear
Security Administration under contract DE-NA0003525. The views expressed in the article
do not necessarily represent the views of the U.S. Department of Energy or the
United States Government. Sandia has major research and development
responsibilities in nuclear deterrence, global security, defense, energy
technologies and economic competitiveness, with main facilities in Albuquerque,
New Mexico, and Livermore, California.


\bibliography{paper}

\begin{thebibliography}{10}
\expandafter\ifx\csname url\endcsname\relax
  \def\url#1{\texttt{#1}}\fi
\expandafter\ifx\csname urlprefix\endcsname\relax\def\urlprefix{URL }\fi
\expandafter\ifx\csname href\endcsname\relax
  \def\href#1#2{#2} \def\path#1{#1}\fi

\bibitem{RAI19}
P.~Rai, K.~Sargsyan, H.~Najm, S.~Hirata, Sparse low rank approximation of
  potential energy surfaces with applications in estimation of anharmonic zero
  point energies and frequencies, Journal of Mathematical Chemistry (2019)
  57:1732--1754.

\bibitem{RAI17}
P.~Rai, K.~Sargsyan, H.~Najm, M.~R. Hermes, S.~Hirata, Low-rank
  canonical-tensor decomposition of potential energy surfaces: application to
  grid-based diagrammatic vibrational green's function theory, Molecular
  Physics 115~(17-18) (2017) 2120--2134.

\bibitem{mezey1}
P.~G. Mezey, Potential energy hypersurfaces.

\bibitem{mezey2}
P.~G. Mezey, Reactive domains of energy hypersurfaces and the stability of
  minimum energy reaction paths, Theoretica chimica acta 54~(2) (1980) 95--111.

\bibitem{mezey3}
P.~G. Mezey, Catchment region partitioning of energy hypersurfaces, i,
  Theoretica chimica acta 58~(4) (1981) 309--330.

\bibitem{mezey4}
P.~G. Mezey, The isoelectronic and isoprotonic energy hypersurface and the
  topology of the nuclear charge space, International Journal of Quantum
  Chemistry 20~(S15) (1981) 279--285.

\bibitem{mezey5}
P.~G. Mezey, Manifold theory of multidimensional potential surfaces,
  International Journal of Quantum Chemistry 20~(S8) (1981) 185--196.

\bibitem{mezey6}
P.~G. Mezey, Critical level topology of energy hypersurfaces, Theoretica
  chimica acta 60~(2) (1981) 97--110.

\bibitem{mezey7}
P.~G. Mezey, The topology of energy hypersurfaces ii. reaction topology in
  euclidean spaces, Theoretica chimica acta 63~(1) (1983) 9--33.

\bibitem{Dawes1}
S.~Manzhos, X.~G. Wang, R.~Dawes, T.~Carrington, A nested molecule-independent
  neural network approach for high-quality potential fits, J. Phys. Chem. A
  110~(16) (2006) 5295--5304.

\bibitem{Dawes2}
R.~Dawes, D.~L. Thompson, Y.~Guo, A.~F. Wagner, M.~Minkoff, Interpolating
  moving least-squares methods for fitting potential energy surfaces: Computing
  high-density potential energy surface data from low-density ab initio data
  points, J. Chem. Phys. 126~(18) (2007) 184108.

\bibitem{Dawes3}
R.~Dawes, D.~L. Thompson, A.~F. Wagner, M.~Minkoff, Interpolating moving
  least-squares methods for fitting potential energy surfaces: A strategy for
  efficient automatic data point placement in high dimensions, J. Chem. Phys.
  128~(8) (2008) 084107.

\bibitem{Nmodeapprox}
S.~Carter, S.~J. Culik, J.~M. Bowman, Vibrational self-consistent field method
  for many-mode systems: A new approach and application to the vibrations of co
  adsorbed on cu(100), J. Chem. Phys. 107~(24) (1997) 10458--10469.

\bibitem{BowmanPES2}
J.~M. Bowman, S.~Carter, X.~C. Huang, Multimode: a code to calculate
  rovibrational energies of polyatomic molecules, Int. Rev. Phys. Chem. 22~(3)
  (2003) 533--549.

\bibitem{BowmanPES}
B.~J. Braams, J.~M. Bowman, Permutationally invariant potential energy surfaces
  in high dimensionality, Int. Rev. Phys. Chem. 28~(4) (2009) 577--606.

\bibitem{BowmanPES3}
J.~M. Bowman, T.~Carrington, H.-D. Meyer, Variational quantum approaches for
  computing vibrational energies of polyatomic molecules, Mol. Phys.
  106~(16-18) (2008) 2145--2182.

\bibitem{Nmodeapprox2}
K.~Yagi, C.~Oyanagi, T.~Taketsugu, K.~Hirao, Ab initio potential energy surface
  for vibrational state calculations of h2co, J. Chem. Phys. 118~(4) (2003)
  1653--1660.

\bibitem{Nmodeapprox3}
K.~Yagi, S.~Hirata, K.~Hirao, Multiresolution potential energy surfaces for
  vibrational state calculations, Theor. Chem. Acc. 118~(3) (2007) 681--691.

\bibitem{Rabitz1}
O.~F. Al{\i}\c{s}, H.~Rabitz, Efficient implementation of high dimensional
  model representations, J. Math. Chem. 29~(2) (2001) 127--142.

\bibitem{Rabitz2}
G.~Y. Li, S.~W. Wang, C.~Rosenthal, H.~Rabitz, High dimensional model
  representations generated from low dimensional data samples. 1. mp-cut-hdmr,
  J. Math. Chem. 30~(1) (2001) 1--30.

\bibitem{Rabitz3}
G.~Y. Li, C.~Rosenthal, H.~Rabitz, High dimensional model representations, J.
  Phys. Chem. A 105~(33) (2001) 7765--7777.

\bibitem{Jackle:1996}
A.~J{\"a}ckle, H.-D. Meyer, Product representation of potential energy
  surfaces, J. Chem. Phys. 104~(20) (1996) 7974--7984.

\bibitem{Jackle:1998}
A.~J{\"a}ckle, H.-D. Meyer, Product representation of potential energy
  surfaces. ii, J. Chem. Phys. 109~(10) (1998) 3772--3779.

\bibitem{Otto:2014}
F.~Otto, Multi-layer potfit: An accurate potential representation for efficient
  high-dimensional quantum dynamics, J. Chem. Phys. 140~(1) (2014) 014106.

\bibitem{Manzhos:2006}
S.~Manzhos, T.~Carrington, Using neural networks to represent potential
  surfaces as sums of products, J. Chem. Phys. 125~(19) (2006) 194105.

\bibitem{Carrington}
G.~Avila, T.~Carrington, Using multi-dimensional smolyak interpolation to make
  a sum-of-products potential, J. Chem. Phys. 143~(4) (2015) 044106.

\bibitem{Rauhut}
B.~Ziegler, G.~Rauhut, Efficient generation of sum-of-products representations
  of high-dimensional potential energy surfaces based on multimode expansions,
  J. Chem. Phys. 144~(11) (2016) 114114.

\bibitem{Rauhut2}
L.~Ostrowski, B.~Ziegler, G.~Rauhut.

\bibitem{hackbusch2008}
W.~Hackbusch, B.~N. Khoromskij, Tensor-product approximation to
  multidimensional integral operators and green's functions, SIAM journal on
  matrix analysis and applications 30~(3) (2008) 1233--1253.

\bibitem{hackbusch2008exp}
W.~Hackbusch, D.~Braess, Approximation of 1/x by exponential sums (2008).

\bibitem{beylkin2009}
G.~Beylkin, C.~Kurcz, L.~Monz{\'o}n, Fast convolution with the free space
  helmholtz green?s function, Journal of Computational Physics 228~(8) (2009)
  2770--2791.

\bibitem{beylkin2005}
G.~Beylkin, L.~Monz{\'o}n, On approximation of functions by exponential sums,
  Applied and Computational Harmonic Analysis 19~(1) (2005) 17--48.

\bibitem{harrison04}
R.~J. Harrison, G.~I. Fann, T.~Yanai, Z.~Gan, G.~Beylkin, Multiresolution
  quantum chemistry: Basic theory and initial applications, The Journal of
  chemical physics 121~(23) (2004) 11587--11598.

\bibitem{khoromskij08}
B.~N. Khoromskij, On tensor approximation of green iterations for kohn-sham
  equations, Computing and visualization in science 11~(4-6) (2008) 259--271.

\bibitem{Grasdyck:2010}
L.~Grasedyck, Hierarchical singular value decomposition of tensors, SIAM J.
  Matrix Anal. Appl. 31~(4) (2010) 2029--2054.

\bibitem{Hackbusch:2009}
W.~Hackbusch, S.~K{\"u}hn, A new scheme for the tensor representation, J.
  Fourier Anal. Appl. 15~(5) (2009) 706--722.

\bibitem{Grasedyck:2013}
L.~Grasedyck, D.~Kressner, C.~Tobler, A literature survey of low-rank tensor
  approximation techniques, GAMM-Mitteilungen 36~(1) (2013) 53--78.

\bibitem{Hermes:2013}
M.~R. Hermes, S.~Hirata, {Second-order many-body perturbation expansions of
  vibrational Dyson self-energies}, J. Chem. Phys. 139~(3) (2013) 034111.

\bibitem{Hermes:2014}
M.~R. Hermes, S.~Hirata, Stochastic many-body perturbation theory for
  anharmonic molecular vibrations, J. Chem. Phys. 141~(8) (2014) 084105, {\it
  ibid.} {\bf 143}, 129903(E) (2015).

\bibitem{Lathauwer00}
L.~De~Lathauwer, B.~De~Moor, J.~Vandewalle, A multilinear singular value
  decomposition, SIAM journal on Matrix Analysis and Applications 21~(4) (2000)
  1253--1278.

\bibitem{kressner2012}
D.~Kressner, C.~Tobler, htucker-a matlab toolbox for tensors in hierarchical
  tucker format, Mathicse, EPF Lausanne.

\bibitem{Oseledets:2011}
I.~V. Oseledets, Tensor-train decomposition, SIAM J. Sci. Comput. 33~(5) (2011)
  2295--2317.

\bibitem{TTToolbox}
I.~Oseledets, S.~DOLGOV, Matlab tt-toolbox version 2.2, 2011.

\bibitem{Hackbusch:2012}
W.~Hackbusch, Tensor Spaces and Numerical Tensor Calculus, Vol.~42, Springer,
  2012.

\bibitem{TTB_Software}
\href{http://www.sandia.gov/~tgkolda/TensorToolbox/}{{\sc MATLAB Tensor Toolbox
  Version 2.6}}, available online, b. W. Bader, T. G. Kolda, {\it et al.}, {\sc
  MATLAB Tensor Toolbox} Version 2.6 (2015), available online at
  http://www.sandia.gov/\~{}tgkolda/TensorToolbox (February 2015).
\newline\urlprefix\url{http://www.sandia.gov/~tgkolda/TensorToolbox/}

\end{thebibliography}

\end{document}